\documentclass[10pt,twocolumn,twoside,conference]{IEEEtran}

\usepackage[pdftex]{graphicx}
\usepackage{cite}

\usepackage[cmex10]{amsmath}
\usepackage{amssymb}
\usepackage{enumerate}
\usepackage{algorithm}
\usepackage{algorithmic}
\usepackage{mdwtab}
\usepackage{amssymb}
\usepackage{eqparbox}
\usepackage{amsfonts}
\usepackage{bm}
\usepackage{epsfig}
\usepackage{booktabs}
\usepackage[caption=false,font=footnotesize]{subfig}
\usepackage{url}
\usepackage{acro}
\usepackage{bbding}
\usepackage{comment}

\usepackage{tikz}
\usepackage{xcolor}
\usetikzlibrary{decorations.pathreplacing}

\usepackage{balance}

\usepackage[utf8]{inputenc}

\newcommand{\Nrf}{N_{\mathrm{RF}}}

\newcommand{\Ntx}{N_{\mathrm{tx}}}
\newcommand{\Lrx}{L_{\mathrm{rx}}}
\newcommand{\Ltx}{L_{\mathrm{tx}}}

\newcommand{\M}[1]{\mathbf{#1}}

\begin{document}

\title{A Geometry-aided Message Passing Method for AoA-Based Short Range MIMO Channel Estimation}
\author{\IEEEauthorblockN{Jarkko Kaleva$^\dagger$, Nitin Jonathan Myers$^*$, Antti T\"olli$^\dagger$ and Robert W. Heath Jr.$^*$%, and Upamanyu Madhow$^\star$
} \\
\IEEEauthorblockA{
    {\it $\dagger$ Centre for Wireless Communications, University of Oulu, Finland.} \\
    {\it $^*$ Department of Electrical and Computer Engineering, The University of Texas at Austin, USA.}\\
    %{\it $^\star$ Department of Electrical and Computer Engineering, University of California at Santa Barbara, USA.}
    }}
\maketitle

\begin{abstract}
Short range channels commonly arise in millimeter wave (mmWave) wearable settings, where the length of the antenna arrays can be comparable to the distance between the radios. Conventional mmWave MIMO channel estimation techniques based on the far field assumption may perform poorly in short range settings due to the large angular spread and, hence, high available rank. We propose a geometry-aided message passing algorithm that exploits structure in short range line-of-sight (LoS) channels for spatial sub-Nyquist channel estimation. Our approach parametrizes the channel using angle-of-arrivals (AoAs) that are locally defined for subarrays of an antenna array. Furthermore, it leverages the dependencies between the local AoAs using factors based on the array geometry. We show that the LoS MIMO channel can be reconstructed using the derived local AoA estimates and the known transceiver geometry. The proposed approach achieves a reasonable rate with greatly reduced pilot transmissions when compared to exhaustive beam search-based local AoA estimation. 
\end{abstract}
%\IEEEpeerreviewmaketitle

\section{Introduction}
Millimeter wave radio architectures can be different from the architectures that are commonly used at lower carrier frequencies~\cite{heathoverview}. The smaller wavelengths at millimeter wave (mmWave) allow the use of a large number of antennas. The number of available radio frequency (RF) chains, however, can be far less than the number of antenna elements, to minimize the power consumption and the cost of mmWave systems~\cite{heathoverview}. The hybrid beamforming architecture, used in the IEEE 802.11ay standard, is one such example. High spectral efficiency can be achieved in a beamforming system if its antenna arrays are configured properly. Such configuration can be achieved if the channel between the transmitter (TX) and the receiver (RX) is known.
\par Channel estimation can be challenging in hybrid beamforming systems due to %the use of large antenna arrays and 
a limited number of RF chains \cite{heathoverview}. To reduce the overhead in learning the channel, prior work has exploited low rank and sparse nature of mmWave channels \cite{javics}. Most of these algorithms, however, make a far field assumption, i.e., the distance between the TX and the RX is larger than the length of the antenna arrays at the TX and the RX \cite{wicomtext}. In typical mmWave wearable settings, the far field assumption may not be applicable. For instance, short range line-of-sight (LoS) channels can have rank that is larger than one; a study on the rank of LoS channels with the transceiver distance can be found in ~\cite{torkildson2011indoor}. In contrast, far field LoS channels have a rank one structure. Similarly, dictionaries used for compressive channel estimation in which far field channels have a sparse representation may not be appropriate in short range settings. Therefore, there is a need to develop algorithms that exploit the short range channel structure to minimize the training overhead for channel estimation.
\par In this paper, we propose a geometry-aided message passing algorithm for short range LoS channel estimation. We consider a point-to-point multiple-input multiple-output (MIMO) communication scenario, where the TX node is equipped with a fully digital architecture and a relatively small number of antennas, while the RX node comprises a large antenna array and a subarray-based hybrid beamforming architecture. Such a setup may be practical in an on-body mmWave sensor network. Assuming channel reciprocity, the estimated channel at the RX node can be used to design the corresponding hybrid beamforming structures for both data transmission and reception.
We assume that the far field assumption holds for the subarray specific channels but not for the full channel. Under such an assumption, we define local angle-of-arrivals (AoAs) for each RX subarray. Our algorithm estimates the local AoAs using the pilots sent from the TX antennas, while exploiting the dependencies among the local AoAs that arise from the geometry of the RX antenna array. The full MIMO channel is then estimated from local AoA information corresponding to two outermost TX antennas of the TX antenna array. The proposed technique has low complexity and performs significantly better than the maximum likelihood approach that recovers local AoAs independently. Simulation results indicate that our algorithm can be used to greatly reduce the training required for LoS channel estimation when compared to conventional techniques.
%\par \textbf{Notation}$:$ $\mathbf{A}$ is a matrix, $\mathbf{a}$ is a column vector and $a, A$ denote scalars. $\mathbf{A}^T$ and $\mathbf{A}^{\ast}$ represent the transpose and conjugate transpose of $\mathbf{A}$. The scalar $a\left[m \right]$ denotes the $m^{\mathrm{th}}$ element of $\mathbf{a}$. $\mathbf{A}(k,:)$ is the $k^{\mathrm{th}}$ row of $\mathbf{A}$ and $\mathbf{A}\left(k,\ell\right)$ is the entry of $\mathbf{A}$ in its $k^{\mathrm{th}}$ row and ${\ell}^{\mathrm{th}}$ column. The Hadamard product of $\mathbf{a}$ and $\mathbf{b}$ is $\mathbf{a} \odot \mathbf{b}$. $\mathcal{I}_N$ denotes the set of integers $\{1,2,...,N \}$. 
\section{System and channel model}
\label{sec:system model}
\begin{figure}[t!]
   \centering
   \includegraphics[width=0.8\columnwidth]{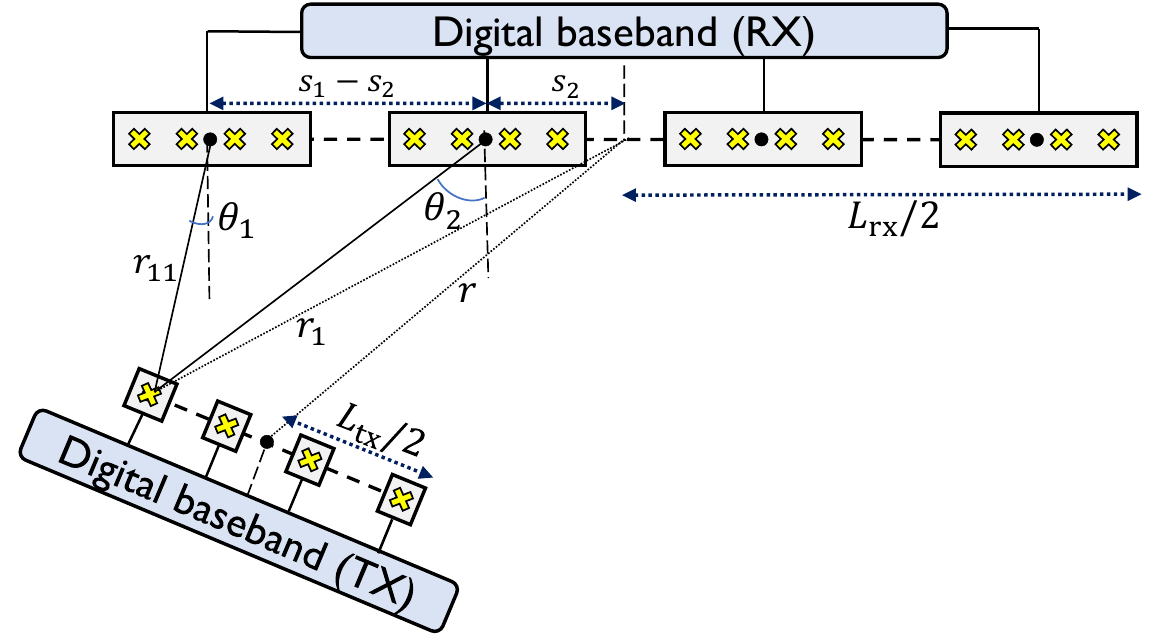}
   \caption{Example of a short range system for $\Nrf=4$, $N=4$, and $N_{\mathrm{tx}}=4$. The angle-of-arrival varies across subarrays when $r$ is comparable to $L_{\mathrm{rx}}$.}
    \label{fig:hybrid}
    \vspace{-0.3cm}
\end{figure}

We consider a point-to-point\footnote{Extension to multiuser scenario is straightforward if each TX is assigned with orthogonal pilots.} MIMO system with a subarray-based hybrid beamforming architecture at the RX illustrated in Fig~\ref{fig:hybrid}. Let $N_{\mathrm{RF}}$ be the number of RF chains and corresponding analog subarrays at the RX. %; the number of RX subarrays is  $N_{\mathrm{RF}}$. 
We use $N_{\mathrm{rx}}$ to denote the number of antennas at the RX. Each of the $\Nrf$ subarrays at the RX is considered to be $\lambda/2$ spaced uniform linear array with $N$ antennas, where $\lambda$ denotes the wavelength corresponding to the mmWave carrier frequency. Furthermore, the $N_{\mathrm{rx}}=N N_{\mathrm{RF}}$ antennas at the RX are assumed to be collinear. Each subarray is equipped with an analog beamforming architecture with phase shifters. The TX is equipped with a fully digital architecture with $\Nrf$ RF chains and $N_{\mathrm{tx}}=\Nrf$ antennas. 
\par We consider a narrowband setting and define $\mathbf{H} \in \mathbb{C}^{N_{\mathrm{rx}} \times N_{\mathrm{tx}}}$ as the channel matrix. The length of the antenna arrays at the TX and the RX are $L_{\mathrm{tx}}$ and $L_{\mathrm{rx}}$, and the distance between the midpoint of the arrays at the TX and the RX is denoted by $r$. 
While the spacing between successive elements in each RX subarray is assumed to be $\lambda/2$, the spacing between the successive subarrays can be arbitrarily larger than $\lambda/2$. We assume that inter-subarray spacing is uniform across successive RX subarrays, and is determined by $\Lrx$ and $N_{\mathrm{RF}}$. Similarly, the spacing between the TX antennas can be larger than $\lambda/2$. The assumptions like the use of uniform subarray spacing, and narrowband setting are made for simplicity of exposition. The ideas underlying our approach, however, can be extended to any TX/RX array geometry.
\par Now, we describe the channel and the system model in the hybrid beamforming setup.  We use $d_{i,j}$ to denote the distance between the $i^{\mathrm{th}}$ antenna at the RX and the $j^{\mathrm{th}}$ antenna at the TX. The $(i,j)^{\mathrm{th}}$ entry of the channel is then \cite{torkildson2011indoor}
\begin{equation}
\label{eq:loschannel}
\mathbf{H}(i,j)= \frac{\lambda}{4\pi d_{i,j}} e^{-\mathsf{j}2 \pi d_{i,j}/ \lambda}.
\end{equation} 
We define $\mathbf{h}_{k,\ell} \in \mathbb{C}^{N}$, a vector in $\mathbf{H}$, as the channel between the $k^{\mathrm{th}}$ subarray at the RX and the ${\ell}^{\mathrm{th}}$ RF chain at the TX.
Let $t_{\ell}[m] \in \mathbb{C}$ be the pilot transmitted by the TX in the $m^{\mathrm{th}}$ training slot. In the same slot, the RX  applies conjugate transpose of $\mathbf{w}_{k}[m] \in \mathbb{C}^N$ to its $k^{\mathrm{th}}$ subarray to acquire channel measurement $y_{k}[m]$. %The conjugate transpose of $\mathbf{w}_{k}[m]$ is defined as $\mathbf{w}^{\ast}_{k}[m]$. 
Under perfect synchronization, the channel measurement is 
\begin{equation}
\label{eq:analog_received_signal}
    y_k[m] = \mathbf{w}^{\ast}_{k} [m] \sum_{\ell = 1}^{N_{\mathrm{RF}}}\mathbf{h}_{k,\ell}t_{\ell}[m] + v_k[m],
\end{equation}
where $v_k[m]\sim \mathcal{N}_{\mathrm{c}}(0, \sigma^2)$ is circularly symmetric Gaussian noise with zero mean and variance $\sigma^2$. As $\mathbf{H}$ has $N \Nrf^2$ entries and the RX can acquire $\Nrf$ channel measurements in parallel, standard channel estimation based on exhaustive search requires a training overhead of $\mathcal{O}(N \Nrf^2/ \Nrf)$. In this paper, we show that, by utilizing the known geometry of the RX antenna array, a reasonable approximation of $\mathbf{H}$ can be estimated with a small fraction of $\mathcal{O}(N \Nrf)$ pilot transmissions.
\section{Geometry-aided channel reconstruction}
%\section{Exploiting structure in short range channels using local AoAs}
The idea underlying the proposed short range LoS channel estimation approach is best understood using local AoA estimation. For ease of exposition, consider a scenario in which the first antenna at the TX transmits known pilots and the rest of the $\Nrf-1$ antennas are inactive. 
For an indicator function $\mathbb{I}$, setting $t_{\ell}[m]=\mathbb{I}_{\ell=1}$ in \eqref{eq:analog_received_signal} results in the desired condition. 
Ignoring the TX antenna index $\ell$, we define $\theta_k$ as the local AoA made by the ray between the midpoints of the $k^{\mathrm{th}}$ RX subarray and the first TX antenna\footnote{In general, $\theta_{k,\ell}$ denotes the local AoA between TX antenna $\ell$ and RX subarray $k$. An identical AoA estimation process is carried out for each $\ell$.}, with the normal to the RX array. %The angle $\theta_k$ is called a local AoA as it is defined for a small segment of the RX array. 
We make an assumption that the AoAs seen by the $N$ individual antennas within subarray $k$ are invariant, and can be approximated by the local AoA $\theta_k$. The angles $\{\theta_k\}^{\Nrf}_{k=1}$, however, can vary with the subarray index $k$, as seen in Fig.~\ref{fig:hybrid}. In mmWave wearable settings, the far field approximation may be valid for each subarray, while it may not hold true for the full array. For example, the length of a virtual reality headset can be comparable to the distance with a wearable. In this paper, we model $\theta_k$ as a realization of a random variable $\Theta_k$, and propose an algorithm to estimate the local AoAs from the  channel measurements while exploiting the dependencies among $\{\theta_k\}^{\Nrf}_{k=1}$. 
\subsection{Construction of likelihood functions for local AoAs} \label{sec:localaoa_ch_meas} %\rev{This seems like AoA estimation is discussed here}
As each subchannel is assumed to satisfy the far field approximation, the local AoAs can be estimated using standard compressed sensing (CS) \cite{csintro}. We use shifted Zadoff-Chu (ZC) sequences as analog RX beamformer weights $\mathbf{w}_k [m]$ for channel acquisition, as ZC sequences can be realized in beamforming systems with unimodulus constraints, and have good properties for AoA estimation \cite{zc_aoa}. For instance, CS-based AoA estimation with ZC sequences is efficient, if the AoAs come from a set of angles that are defined by the discrete Fourier transform. We define $\mathbf{z}_k \in \mathbb{C}^N$ as the ZC sequence used at the $k^{\mathrm{th}}$ subarray of the RX. Spatially diverse channel measurements are acquired at the RX by applying different random circulant shifts of $\mathbf{z}_k$ to its $k^{\mathrm{th}}$ subarray.
\par For pilot transmission from the first antenna at the TX, the $m^{\mathrm{th}}$ channel measurement at the $k^{\mathrm{th}}$ RX subarray is  
\begin{equation}
y_k[m]=\mathbf{w}^{\ast}_k [m] \mathbf{h}_{k,1} +v_k[m].
\end{equation}
We define a Vandermonde vector of length $N$ as $\mathbf{a}_N(\theta)=[1,e^{-\mathsf{j}\pi \mathrm{sin}(\theta)},e^{-\mathsf{j}2\pi\mathrm{sin}(\theta)},..,e^{-\mathsf{j}(N-1)\pi \mathrm{sin}(\theta)}]^T$. Under the far field assumption for subchannels, $\mathbf{h}_{k,1}$ can be approximated as $\alpha_k \mathbf{a}_N(\theta_k)$, where $\alpha_k$ is an unknown complex gain. The channel measurement $y_k[m]$ is then
\begin{equation}
\label{eq:y_k_m_vand}
y_k[m]=\alpha_k \mathbf{w}^{\ast}_k [m] \mathbf{a}_N(\theta_k) + v_k[m].
\end{equation} 
A collection of $M$ projections of $\mathbf{a}_N(\theta_k)$, obtained using RX beam training vectors $\{ \mathbf{w}_k [m] \}^M_{m=1}$, is defined as $\mathbf{y}_k \in \mathbb{C} ^{M}$. In this paper, the first $M-1$ beam training vectors, i.e., $\{ \mathbf{w}_k [m] \}^{M-1}_{m=1}$, are chosen as $M-1$ distinct random circulant shifts of $\mathbf{z}_k$. The vector $\mathbf{w}_{k}[M]$ is defined as $\mathbf{w}_{k}[M]=\mathbf{w}_{k}[1]\odot [1,-1,-1,..-1]^T$, where $\odot$ denotes the element-wise product. The $M^{\mathrm{th}}$ measurement is defined differently so that the unknown gain $\alpha_k$ can be estimated. With $\eta_k$ defined as the first entry of $\mathbf{w}_{k}[1]$, it can be observed from \eqref{eq:y_k_m_vand} that $y_k[1]+y_k[M]$ is a noisy version of $2 \alpha_k \eta_k$. An estimate of $\alpha_k$ is then $\hat{\alpha}_k=(y_k[1]+y_k[M])/2\eta_k$.   
\par The gain compensated channel measurements are defined as $\tilde{\mathbf{y}}_k = \mathbf{y}_k/\hat{\alpha}_k$. The compression matrix associated with the $k^{\mathrm{th}}$ subarray is defined as $\mathbf{A}_k \in \mathbb{C}^{M \times N}$, where the $m^{\mathrm{th}}$ row of $\mathbf{A}_k$ is $\mathbf{A}_k(m,:)=\mathbf{w}^{\ast}_{k}[m]$. From \eqref{eq:y_k_m_vand}, it can be observed that $\mathbf{y}_k = \alpha_k \mathbf{A}_k \mathbf{a}_N(\theta_k) + \mathbf{v}_k$. We ignore the errors in estimating $\alpha_k$ to conclude that $\tilde{\mathbf{y}}_k$ is a realization of $\mathcal{N}_{\mathrm{c}}(\mathbf{A}_k \mathbf{a}_N(\theta_k), \sigma^2 \mathbf{I}/|\hat{\alpha}_k|^2)$. Thus, the scaled likelihood function $p(\theta_k)$ is defined as 
\begin{equation}
\label{eq:likelihood}
p(\theta_k)=\mathrm{exp}\left(-{|\hat{\alpha}_k|^2 \Vert \tilde{\mathbf{y}}_k - \mathbf{A}_k \mathbf{a}_N(\theta_k) \Vert^2}/{\sigma^2}\right).
\end{equation}    
A possible way to estimate $\theta_k$ is by maximizing $p(\theta_k)$ in \eqref{eq:likelihood}. 
\subsection{Statistical dependency among local AoAs} 
In this section, we design geometry factors to model the strong dependencies among the local AoAs $\{\theta_k\}^{\Nrf}_{k=1}$. For a specific $\theta_1$, it can be observed from Fig.~\ref{fig:hybrid} that $\theta_2$ is a function of the distance between the first antenna at the TX and the first subarray of the RX. As this distance is typically bounded, $\theta_2$ takes a range of values that depend on $\theta_1$. The maximum likelihood (ML) estimator that independently maximizes \eqref{eq:likelihood} over $\{\theta_k\}_{k=1}^{\Nrf}$ does not exploit such dependency, and may result in poor local AoA estimates. 
\par We derive the geometry factor $g(\theta_2|\theta_1)$ that represents the distribution of $\Theta_2$ conditioned on $\Theta_1= \theta_1$. We define $r_1$ as the distance between the first TX antenna  and the midpoint of the RX array, and $s_k$ as the distance of the $k^{\mathrm{th}}$ RX subarray from the midpoint of the RX. For a given $r_1$, the distance $r_{11}$ in Fig.~\ref{fig:hybrid} can be solved from
\begin{equation}
\label{eq:r11}
(r_{11}\mathrm{cos}\,\theta_{1})^{2}+(r_{11}\mathrm{sin}\,\theta_{1}+s_{1})^{2}=r_{1}^{2}
\end{equation} 
and, subsequently, the local AoA $\theta_2$ can be expressed as 
\begin{equation}
\label{eq:theta2solve}
\theta_2=\mathrm{tan}^{-1}\left(\frac{s_1-s_2+r_{11}\mathrm{sin}\,\theta_{1}}{r_{11}\mathrm{cos}\,\theta_{1}}\right).
\end{equation}
From \eqref{eq:r11} and \eqref{eq:theta2solve}, it can be observed that there is a unique $\theta_2$ for a given $\theta_1$ and $r_1$. Let $\mathcal{G}$ be the mapping from $\theta_1$ and $r_1$ to $\theta_2$, i.e., $\theta_2=\mathcal{G}(\theta_1,r_1)$. We assume that $r\in [r_{\mathrm{min}},r_{\mathrm{max}}]$. For example, it is reasonable to assume $r_{\mathrm{min}} \approx 15 \, \mathrm{cm}$ and $r_{\mathrm{max}} \approx 80 \, \mathrm{cm}$ in on-body communication setups. The distance $r_1$ is known to lie in $[r_{\mathrm{min}}-\Ltx/2, r_{\mathrm{max}}+\Ltx/2]$. We assume that $r_1$ is uniformly distributed within this interval to get
\begin{equation}
\label{eq:geo_fact}
g(\theta_{2}|\theta_{1})=\frac{1}{D}\int_{r_{\mathrm{min}}-\Ltx/2}^{r_{\mathrm{max}}+\Ltx/2}\delta(\theta_2-\mathcal{G}(\theta_{1},r_{1}))dr_{1},
\end{equation}  
where $D=r_{\mathrm{max}}-r_{\mathrm{min}}+\Ltx$ and $\delta(\cdot)$ denotes the Dirac-delta function indicating $\theta_2=\mathcal{G}(\theta_1,r_1)$. Similarly, other conditional distributions, i.e., $\{g(\theta_k|\theta_n)\}_{k,n}$, can be estimated with the arguments used to compute $g(\theta_2|\theta_1)$. The geometry factors do not depend on the channel measurements, and can be computed offline based on the array geometry at the RX.
\subsection{Geometry-aided message passing}\label{sec:geom_mp}
\begin{figure}[t!]
   \centering
   \includegraphics[width=0.8\columnwidth]{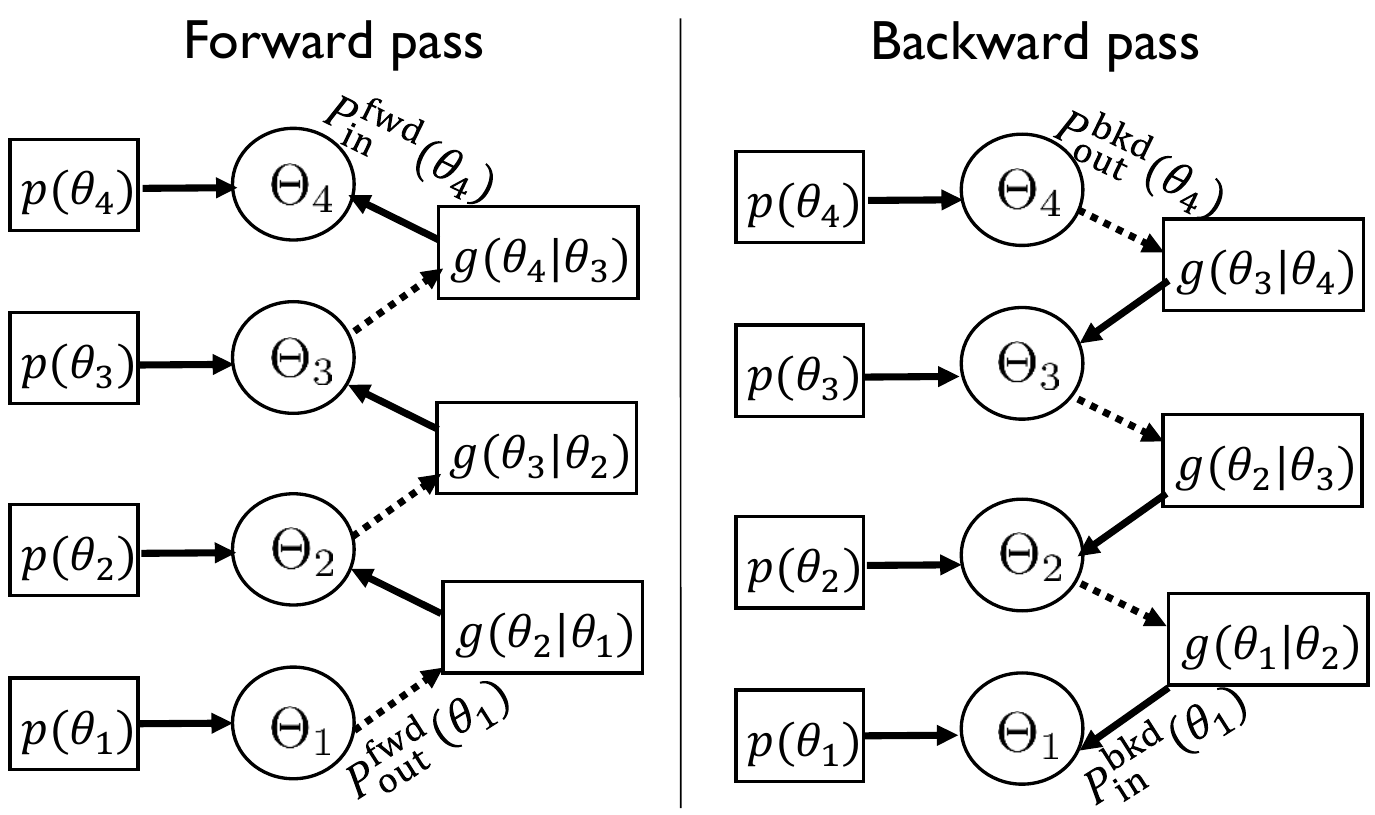}
   \caption{Factor graphs in forward and backward passes of message passing for $\Nrf=4$. The random variable $\Theta_k$ models a realization of $\theta_k$.}
   \label{fig:fact}
   \vspace{-0.3cm}
\end{figure}
Now, we explain our algorithm that combines information about $\{\Theta_k\}_{k=1}^{\Nrf}$ from the likelihood functions~\eqref{eq:likelihood} and the geometry factors~\eqref{eq:geo_fact}. The factor graph~\cite{SPA} corresponding to our algorithm is shown in Fig.~\ref{fig:fact}. The circular nodes in Fig.~\ref{fig:fact} denote random variables, and the rectangular nodes contain factors or likelihoods. A message flowing out of a node represents a distribution of a random variable, from the node's perspective. Messages are sent between the circular nodes and the rectangular nodes to obtain better estimates of $\{\theta_k\}_{k=1}^{\Nrf}$ when compared to the ML estimator. Our algorithm includes a forward pass, i.e., sequential flow of messages among $\{\Theta_k\}^{\Nrf}_{k=1}$, and a backward pass in which message flows occur in the opposite direction. The messages in both directions are computed using the sum-product algorithm~\cite{SPA}.
\par We explain how messages are constructed in the forward pass. In the first iteration of the forward pass, the likelihood function $p(\theta_1)$ is sent to node $\Theta_1$, which forwards $P^{\mathrm{fwd}}_{\mathrm{out}}(\theta_{1})=p(\theta_1)$ to the geometry factor that contains $g(\theta_2|\theta_1)$. %We define $P^{\mathrm{fwd}}_{\mathrm{out}}(\theta_{1})=p(\theta_1)$. %At this point, the geometry factor has the outflow from node $\Theta_1$, i.e., $P^{\mathrm{fwd}}_{\mathrm{out}}(\theta_{1})$, and contains the conditional distribution $g(\theta_2|\theta_1)$. 
Using the two functions, the geometry factor forwards message $P^{\mathrm{fwd}}_{\mathrm{in}}(\theta_{2})$ to node $\Theta_2$, defined by
\begin{equation}
\label{eq:pfwd_in_theta2}
P^{\mathrm{fwd}}_{\mathrm{in}}(\theta_{2})=\int_{-\pi/2}^{\pi/2}P^{\mathrm{fwd}}_{\mathrm{out}}(\theta_{1})g(\theta_{2}|\theta_{1})d\theta_{1}.
\end{equation}
The distribution $P^{\mathrm{fwd}}_{\mathrm{in}}(\theta_{2})$ is essentially $P^{\mathrm{fwd}}_{\mathrm{out}}(\theta_{1})g(\theta_{2}|\theta_{1})$ marginalized over $\theta_1$. In other words, $P^{\mathrm{fwd}}_{\mathrm{in}}(\theta_{2})$ represents the belief about $\Theta_2$, i.e., a scaled probability distribution of $\Theta_2$ believed by the geometry factor, using information about $\Theta_1$. Note that $P^{\mathrm{fwd}}_{\mathrm{in}}(\theta_{2})$ provides side information about $\Theta_2$ that is independent of the measurements acquired by the second subarray. This side information comes from the observations in the first subarray, and the statistical dependency between $\theta_1$ and $\theta_2$. The node $\Theta_2$ combines information from the channel measurements, i.e., $p(\theta_2)$, with the one from $P^{\mathrm{fwd}}_{\mathrm{in}}(\theta_{2})$, using 
\begin{equation}
\label{eq:pfwd_out_theta2}
P^{\mathrm{fwd}}_{\mathrm{out}}(\theta_2)=p(\theta_2)P^{\mathrm{fwd}}_{\mathrm{in}}(\theta_{2}).
\end{equation} 
The message in \eqref{eq:pfwd_out_theta2} is sent to the factor containing $g(\theta_3|\theta_2)$, which computes the belief about $\Theta_3$ with $g(\theta_3|\theta_2)$ and $P^{\mathrm{fwd}}_{\mathrm{out}}(\theta_2)$. %The message sent by the factor, i.e., $P^{\mathrm{fwd}}_{\mathrm{in}}(\theta_{3})$, is computed using an expression similar to \eqref{eq:pfwd_in_theta2}. 
The process of message flows continues until the last node with $\Theta_{\Nrf}$ is reached. The messages in \eqref{eq:pfwd_in_theta2} and \eqref{eq:pfwd_out_theta2} can be generalized, by setting $\theta_1$ to $\theta_{k-1}$ and $\theta_{2}$ to $\theta_{k}$, to obtain recursive equations. The forward pass computes message inflows $\{P^{\mathrm{fwd}}_{\mathrm{in}}(\theta_{k})\}_{k=2}^{\Nrf}$ using the recursive equations.  
\par The forward pass does not exploit information about $\{\theta_n\}^{\Nrf}_{n=k}$ to generate side information about $\theta_{k-1}$. The backward pass overcomes this issue by performing message flows in the opposite direction of the forward pass. In the first iteration, message from the $\Nrf^\text{th}$ node, i.e., $P^{\mathrm{bwd}}_{\mathrm{out}}(\theta_{\Nrf})=p(\theta_{\Nrf})$, flows into the geometry factor containing $g(\theta_{\Nrf-1}|\theta_{\Nrf})$. The geometry factor then sends a belief about $\Theta_{\Nrf-1}$, defined as $P^{\mathrm{bwd}}_{\mathrm{in}}(\theta_{\Nrf-1})$, based on $p(\theta_{\Nrf})$. Similarly, the backward pass is computed using    
\begin{align}
\label{eq:pbwd_in_thetak}
P^{\mathrm{bwd}}_{\mathrm{in}}(\theta_{k-1})&=\int_{-\pi/2}^{\pi/2}P^{\mathrm{bwd}}_{\mathrm{out}}(\theta_{k})g(\theta_{k-1}|\theta_{k})d\theta_{k},\\
P^{\mathrm{bwd}}_{\mathrm{out}}(\theta_{k})&=p(\theta_k)P^{\mathrm{bwd}}_{\mathrm{in}}(\theta_{k}).
\end{align}
It can be observed that nodes $\Theta_1$ and $\Theta_{\Nrf}$ do not receive side information in the forward and backward passes. Therefore, we set $P^{\mathrm{fwd}}_{\mathrm{in}}(\theta_{1})=\mathcal{U}(\theta_1)$ and $P^{\mathrm{bwd}}_{\mathrm{in}}(\theta_{\Nrf})=\mathcal{U}(\theta_{\Nrf})$, where $\mathcal{U}(\theta)$ denotes a uniform distribution over $\theta$.   
\par At the end of forward and backward passes, each node $k$ obtains side information in the form of $P^{\mathrm{fwd}}_{\mathrm{in}}(\theta_{k})$ and $P^{\mathrm{bwd}}_{\mathrm{in}}(\theta_{k})$, and also has access to the likelihood $p(\theta_k)$. The three sources of information about $\Theta_k$ can be combined by defining a new distribution $p_{\mathrm{gmp}}(\theta_k)$, i.e., 
\begin{equation}
\label{eq:combprob}
p_{\mathrm{gmp}}(\theta_k)=p(\theta_k)P^{\mathrm{fwd}}_{\mathrm{in}}(\theta_{k})P^{\mathrm{bwd}}_{\mathrm{in}}(\theta_{k}).
\end{equation}
Finally, the local AoA estimate with our geometry-aided message passing algorithm is defined as $\hat{\theta}_k=\mathrm{arg\,max}\,p_{\mathrm{gmp}}(\theta_k)$. For computational tractability, the integrals in \eqref{eq:geo_fact}, \eqref{eq:pfwd_in_theta2}, and \eqref{eq:pbwd_in_thetak} are computed using a discrete sum, assuming angular resolution of $\kappa \pi$. Thus, the complexity of the algorithm is $\mathcal{O}(\Nrf/\kappa^2)$. It can be noticed that the factor graph in our algorithm models dependencies among adjacent local AoAs, and does not include factors like $g(\theta_k|\theta_n)$ for $|k-n|>1$. Incorporating such factors can result in short cycles, which may not be desirable from a message passing perspective \cite{cycles_avoid}.
\subsection{MIMO channel reconstruction from local AoA estimates}
A reasonable approximation of the LoS MIMO channel matrix can be estimated at the RX from the local AoA estimates. First, we note that the RX knows the geometry of its antenna array, and can compute the position vectors of all the RX antenna elements relative to a common reference. Let $\mathbf{s}_1$ and $\mathbf{s}_{\Nrf}$ denote the position vectors of the first and $\Nrf^{\mathrm{th}}$ RX subarrays. We define $\mathbf{s}=(\mathbf{s}_{1}-\mathbf{s}_{\Nrf})/\Vert \mathbf{s}_{1}-\mathbf{s}_{\Nrf}\Vert_2$ as a unit vector along the antenna array at the RX. It can be observed from Fig. \ref{fig:hybrid} that the first TX antenna is at a distance of $r_{11}$ from the first RX subarray. Furthermore, the direction of this TX antenna, relative to the RX array, can be found by rotating $\mathbf{s}$ in the anti-clockwise direction by $\pi/2-\theta_1$. For a rotation matrix $\mathbf{R}(\Delta)$ given by 
\begin{equation}
\label{eq:rotation_1}
    \mathbf{R}(\Delta) = 
    \begin{bmatrix}
        \cos(\Delta) & -\sin(\Delta) \\
        \sin(\Delta) & \cos(\Delta)
    \end{bmatrix}
    \text{,}
\end{equation}
the position vector of the first TX antenna is defined as 
\begin{math}
%\label{eq:pv_tx_1}
\mathbf{q}_1= \mathbf{s}_{1}+r_{11}\mathbf{R}(\pi/2-\theta_1)\mathbf{s},
\end{math} 
where $r_{11}$ can be computed using triangulation from multiple local AoA estimates. %In this paper, however, we approximate $r_{11}$ simply with $(r_{\mathrm{min}}+r_{\mathrm{max}})/2$. 
The position vector of the $\Nrf^{\mathrm{th}}$ TX antennas can be derived by the algorithm in Section~\ref{sec:geom_mp} to estimate the local AoAs corresponding to the $\Nrf^{\mathrm{th}}$ TX. Finally, the  coordinates of the remaining TX antennas are acquired using the estimated positions of the first and last TX antennas and the known TX geometry. The LoS channel is constructed from the estimated coordinates using~\eqref{eq:loschannel}. Note that the pilot transmissions can be performed from different antennas at the TX using spread sequences.
\section{Simulation Results}
We consider $N=16$ antennas per subarray, and $\Nrf=4$ subarrays for the hybrid beamforming architecture at the RX. The TX is equipped with a fully digital architecture with $\Ntx=4$ antennas and $\Nrf=4$ RF chains. The carrier frequency in the system is set to $60\, \mathrm{GHz}$, which corresponds to $\lambda=5\, \mathrm{mm}$. The length of the arrays at the TX and the RX are $\Ltx=4\, \mathrm{cm}$ and $\Lrx=18\,\mathrm{cm}$. It may be possible to mount the RX array on a augmented reality headset, and the TX array on a wearable like smart watch. The spacing between the midpoint of successive RX subarrays is $4.75\, \mathrm{cm}$, and the spacing between neighbouring TX antennas is $1.33\, \mathrm{cm}$. 
The TX and RX arrays are placed on a horizontal plane at a height of $1.5\, \mathrm{m}$, in a room of dimensions $5\, \mathrm{m} \times 5\, \mathrm{m} \times 3\, \mathrm{m}$. The properties of the ceiling, side-walls, and the floor  were set according to the parameters in \cite{torkildson2011indoor}. The channel in our simulations has components corresponding to reflections from the walls in addition to the LoS component. Our algorithm, however, can only estimate the dominant signal path corresponding to the LoS component. The distance between the transceivers, i.e., $r$, is chosen uniformly at random from $[r_{\mathrm{min}},r_{\mathrm{max}}]$ for $r_{\mathrm{min}}=40\, \mathrm{cm}$ and $r_{\mathrm{max}}=80\, \mathrm{cm}$. The orientations of the TX and RX arrays are chosen at random to cover a wide range of possible configurations. The RX SNR for pilot measurements, $5\, \mathrm{dB}$, is chosen to be fairly small to demonstrate the resilience of the algorithm against estimation inaccuracies.  
\begin{figure}[t!]
    \centering
    \includegraphics[trim=1.5cm 7.25cm 2cm 7.4cm,clip=true,width=0.45 \textwidth]{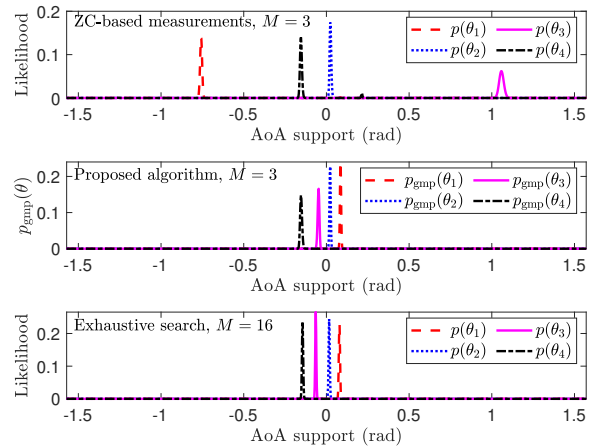}
    \caption{ Likelihood functions $\{p (\theta_k)\}_{k=1}^{\Nrf}$ for a particular channel realization. %The estimated distributions, i.e., $\{p_{\mathrm{gmp}} (\theta_k)\}_{k=1}^{\Nrf}$, achieve maximum near the directions obtained with exhaustive scan.
    }
    \label{fig:aoa}
    \vspace{-0.5cm}
\end{figure}
\par
\begin{figure}[t!]
    \centering
    \includegraphics[trim=1.8cm 6.25cm 2cm 7.4cm,clip=true,width=0.85 \columnwidth]{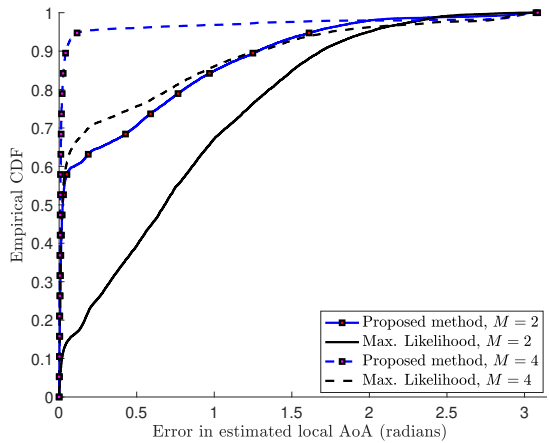}
    \vspace{-0.3cm}
    \caption{The proposed geometry-aided message passing approach results in small local AoA errors when compared to the maximum likelihood approach.}
    \label{fig:cdf}
    \vspace{-0.3cm}
\end{figure}

 We evaluate three different algorithms in a short range setting. The first method uses the ZC-based compression matrix defined in Sec. \ref{sec:localaoa_ch_meas}, and independently maximizes the likelihood functions in \eqref{eq:likelihood}. The second approach estimates local AoAs using the proposed geometry-aided message passing algorithm. Our algorithm uses information from the likelihoods, and the geometry factors that are based on the RX array configuration. For a benchmark, we define the third approach based on  exhaustive beam search, i.e., the DFT dictionary is used for $\mathbf{A}_k \ \forall \ k$, and $M=16$ channel measurements are acquired in each subarray. The resolution in the angle ($\theta$) space is set to $=0.125^{\mathrm{o}}$, using $\kappa=1/1440$. An example of the likelihood functions $\{p(\theta_k)\}_{k=1}^{\Nrf}$ is shown in Fig.~\ref{fig:aoa}. % for $M=3$ ZC-based channel measurements per subarray. 
It can be observed %from Fig.~\ref{fig:aoa}
that maximizing the likelihood functions independently for $M=3$ can result in local AoAs that are significantly different from each other. The proposed method, however, is able to ``repair'' the mismatched AoAs using geometry factors and message passing. In Fig.~\ref{fig:cdf}, we plot the empirical cumulative distribution function (ECDF) of the errors in the local AoA estimates with the maximum likelihood approach and the proposed method. It can be observed from Fig.~\ref{fig:cdf} that the local AoAs recovered by our algorithm are significantly closer to those obtained with the benchmark, when compared to the maximum likelihood approach. 
\begin{figure}[t]
    \centering
    \includegraphics[trim=2cm 6.25cm 2cm 7.4cm,clip=true,width=0.85 \columnwidth]{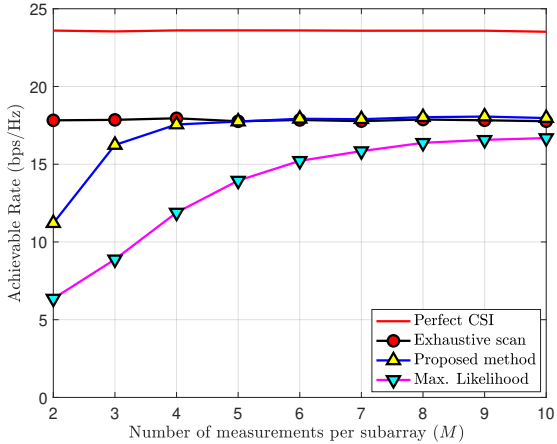}
    \vspace{-0.3cm}
    \caption{Short range channels can be estimated with fewer pilot transmissions by exploiting local angle dependencies that arise from the array geometry.}
    \label{fig:mse_M}
    \vspace{-0.4cm}
\end{figure}
\balance % This should be placed on the left column of the last page
\par The accuracy of the reconstructed MIMO channel is studied using the achievable rate obtained with the estimated channel as a figure of merit. The rate analysis with practical hybrid beamforming structures is left for future study. Let $\tilde{\M{H}} = \tilde{\M{U}}\tilde{\M{D}}\tilde{\M{V}}^\ast$ denote the singular value decomposition (SVD) of the estimated channel. Singular matrices $\tilde{\M{U}}$ and $\tilde{\M{V}}$ are considered to configure the precoders and the combiners at the TX and the RX. The achievable rate of the system (assuming ideal all-digital RX) is determined from the capacity of MIMO channel $\tilde{\M{U}}^\ast \M{H} \tilde{\M{V}}$. The simulation results are then averaged over $2500$ random orientations. It can be observed from Fig.~\ref{fig:mse_M} that channel estimation using the proposed algorithm results in a reasonable achievable rate that is comparable to the one achieved with exhaustive scan. While the exhaustive scan requires $16$ channel measurements per subarray, our algorithm achieves good performance with just $M=3$ measurements per subarray. The performance gap to the perfect CSI case is due to fairly low $5$dB SNR available for estimation of both local AoA $\theta_{k,\ell}$ and $\alpha_k$ in~\eqref{eq:y_k_m_vand}, as well as, unaccounted non-LoS components in the actual channel model.
\section{Conclusions}
Short range channels exhibit different structure when compared to the commonly studied far field channels. For example, the angles-of-arrival in a short range setting can vary across multiple sections of the receive antenna array. In this paper, we have shown that the angles can be statistically dependent, with the dependencies determined by the receive array geometry. We have developed a message passing algorithm that exploits the statistical dependencies between the angles, for channel estimation with sub-Nyquist measurements in the angular domain. Our results indicate that geometry information can serve as strong regularizer for the channel estimation problem.

%\vspace{-2px}
%% Acknowledgements ---------------------------------------
%\small
\section*{Acknowledgements}
%This research is supported in part by the Academy of Finland under grant number 318927, and by the U.S. National Science Foundation under grant number NSF-CNS-1702800.
This research is supported by the Academy of Finland under grant numbers 311741 and 318927 (6Genesis Flagship), and by the U.S. National Science Foundation under grant numbers CNS-1702800 and ECCS-1711702.
%projects: WiFiUS: Millimeter Wave-based Wearable Networks in High-end IoT Applications (grant 311741) and 6Genesis Flagship (grant 318927). 
%\normalsize

\bibliographystyle{IEEEtran}
\bibliography{refs}

\end{document}